\documentclass[aps,twocolumn,showpacs,groupedaddress]{revtex4}  % for review and submission
\pdfoutput=1 
\usepackage{graphicx}  % needed for figures
\usepackage{subfigure} % needed for subfigures 
\usepackage{dcolumn}   % needed for some tables
\usepackage{bm}        % for math
\usepackage{amssymb}   % for math
\usepackage{amsmath}	   % for math
\usepackage{lipsum}		% for long equations 

\begin{document}
\title{Non-equilibrium Quantum Langevin dynamics of orbital diamagnetic moment}
\author{Urbashi Satpathi}
\address{Raman Research Institute, C. V. Raman Avenue,
Sadashivanagar, Bangalore 560080, India.}
\author{Supurna Sinha}
\address{Raman Research Institute, C. V. Raman Avenue,
Sadashivanagar, Bangalore 560080, India.}
%*\input author_list.tex       % D0 authors (remove the first 3 lines
                             % of this file prior to submission, they
                             % contain a time stamp for the authorlist)
                             % (includes institutions and visitors)
\date{\today}
\begin{abstract}

We investigate the time dependent orbital diamagnetic moment of a charged particle 
in a magnetic field in a viscous medium via the 
Quantum Langevin Equation. We study how the interplay between the cyclotron frequency and the viscous damping rate governs the dynamics of the orbital magnetic moment
in the high temperature classical domain and  the low temperature quantum domain for 
an Ohmic bath. These predictions can be tested via state of the art cold atom experiments with hybrid traps for ions and neutral atoms. We also study the effect of a confining potential on the dynamics of the magnetic moment. We obtain the expected Bohr Van Leeuwen limit in the high temperature, asymptotic time ($ \gamma t\longrightarrow \infty$, where $ \gamma $ is the viscous damping coefficient) limit.

\end{abstract}

\pacs{}
\maketitle
\section{Introduction}
The problem of a quantum charged particle in the presence of a magnetic field, coupled to a quantum mechanical 
heat bath has been the focus of
many studies\cite{mexicanpaper,ford2,cob,shamik,Dattaguptaprl,marathe,paper2}. The paradigm that is commonly used is the Quantum
Langevin Equation\cite{ford,ford2} where the quantum charged particle is linearly coupled to  
a heat bath. The independent oscillator (IO) model is typically used for the heat bath. The coupling with the heat bath is characterized by 
the operator valued random force $F(t)$ and a memory function $\mu(t)$ which corresponds to the dissipative kernel.  
The Quantum Langevin Equation is obtained
by integrating out the bath degrees of freedom, via the Heisenberg description\cite{ford,ford2}.  

Our interest here is to study the time dependent orbital magnetic moment using the nonequilibrium Quantum Langevin dynamics. 
While there has been quite a lot of interest in this area,
most of the studies have focused on the asymptotic limit of time $t\rightarrow\infty$ \cite{marathe,Dattaguptaprl}. 
Here we investigate in detail the time dependent orbital magnetic moment and study the interplay between the
effect of the magnetic field and damping effects due to dissipation. In particular, a particle of
charge $q$ and mass $m$
in a magnetic field $B$, moves in a circular orbit at a rate set by the cyclotron frequency $\omega_c= qB/{mc}$, where
$c$ is the speed of light.
The friction coefficient $\gamma$ provides a rate of dissipation. We probe various different
regimes of these two competing time scales both in the high temperature classical domain and the low temperature quantum
domain and analyze the dynamics of the orbital magnetic moment in these regimes.

We notice that the presence of these competing time scales bring about some subtleties in determining the asymptotic time behaviour $ - $ in particular the high temperature Bohr Van Leeuwen limit of a zero magnetic moment.
In earlier studies\cite{Dattaguptaprl,marathe} a confining potential has been introduced in order to recover the 
equilibrium Landau diamagnetism expression for the magnetic moment, which in turn, leads to 
the Bohr Van Leeuwen limit for the magnetic moment in the classical high temperature domain. However, the final expression for the Landau diamagnetic moment has been obtained in the limit of a zero confining potential. In this paper, we go beyond earlier studies in investigating in detail the dependence of the magnetic moment on the confining potential. In fact the confining potential which is of the harmonic oscillator potential form $ V=\frac{1}{2}m\omega_0^{2}r^2 $, brings in an additional time scale $ \omega_{0}^{-1} $ into the problem.
This enables us to study the relative importance of the time scales set by the viscous damping rate $\gamma$, the cyclotron frequency $ \omega_c $ and the frequency $ \omega_{0} $ of the harmonic potential and study the emergence of a zero magnetic moment in the long time,
high temperature limit. In contrast to earlier studies we notice that the Bohr Van Leeuwen zero magnetic
moment limit can be obtained even without introducing an external potential. Needless to say, one 
does recover the Bohr Van Leeuwen limit also in the presence of a potential.

This paper is organized as follows. In Sec II we solve the Quantum Langevin equation for a charged particle in an external potential and in the presence of a magnetic field and analyse the general solution in terms of the retarded Green's function. In Sec III we study the magnetic moment in the absence of a potential for both high temperature classical and low temperature quantum domains. In both cases, we analyse two regimes-a viscosity dominated regime and a magnetic field dominated regime. In Sec IV we do a similar analysis in the presence of a potential. In Sec V we study the approach to a high temperature classical thermodynamic limit both-in the absence and in the presence of a potential. Finally, we end our paper with some concluding remarks in Sec VI.    

\section{Quantum Langevin equation for a charged particle in a magnetic field}
The Quantum Langevin Equation (QLE) of a charged particle in an external potential and in the presence of a magnetic field is given by \cite{ford2}
\begin{eqnarray}
m \ddot{\vec{r}}(t)&=&-\int_{0}^{t}\mu (t-t')\dot{\vec{r}}(t')dt'+\frac{q}{c}(\dot{\vec{r}}(t)\times\vec{B})\nonumber\\
&&-V'(\vec{r}(t))+\vec{f}(t)\label{e1}
\end{eqnarray}
where, $ m $ is the mass of the particle, $ \mu(t) $ is the memory kernel, $ q $ is the charge, $ c $ is the speed of light, $ \vec{B} $ is the applied magnetic field, $ V $ is an external potential and $ \vec{f}(t) $ is the random force operator. The QLE is a Heisenberg equation of motion for the displacement variable $ \vec{r}(t) $. The quantum particle is linearly coupled to a passive heat bath. The couplings with the heat bath are described by the random force operator and the memory kernel.
The spectral properties of the random force operator are characterized by the symmetric correlation and the commutator:
\begin{eqnarray}
\frac{1}{2}\langle \left\lbrace f_{\alpha}(t),f_{\beta}(0) \right\rbrace \rangle &=& \frac{\delta_{\alpha\beta}}{2\pi }\int_{-\infty}^{\infty}d \omega \mathrm{Re}\left[ \mu(\omega)\right]\hbar\omega \nonumber\\
&&\mathrm{coth}\left(\frac{\hbar\omega}{2k_{B}T} \right) e^{-i\omega t}\label{e3}\\
\langle\left[ f_{\alpha}(t),f_{\beta}(0)\right]\rangle &=& \frac{\delta_{\alpha\beta}}{\pi}\int_{-\infty}^{\infty}d \omega \mathrm{Re}\left[ \mu(\omega)\right] \nonumber\\
&&\hbar\omega e^{-i\omega t}\label{e4}
\end{eqnarray}
Here, $ \alpha,\beta=x,y,z $, and $ \delta_{\alpha\beta} $ is the Kronecker delta function. 
%Note that the above equations are direct consequence of Fluctuation dissipation theorem. The fluctuation dissipation theorem relates the fluctuating and the dissipative parts of the QLE, which are the random force and the memory kernel, respectively.

Let us consider the magnetic field to be along the $ z $ axis. 
Then the motion of the particle in the $ x-y $ plane is affected by the presence of the magnetic field. The motion along the $ z $ axis is a free particle motion. 
The motion in the $ x-y $ plane is given by:
\begin{eqnarray}
\ddot{\xi}(t)&=&-\int_{0}^{t}K(t-t')\dot{\xi}(t')dt'-\frac{iqB}{mc}\dot{\xi}(t)-\omega_{0}^{2}\xi(t)\nonumber\\
&+&\frac{F(t)}{m}\label{q1}
\end{eqnarray}
where, $ K(t)=\frac{\mu(t)}{m} $.
In the above equation we consider $ V(x,y) = \frac{1}{2}k(x^2+y^2) $. Here,
\begin{eqnarray}
 \xi=x+iy , \;\; 	\omega_{0}=\sqrt{\frac{k}{m}}, \;\;	F = f_{x}+if_{y} \label{q2}
\end{eqnarray} 
%In the next section keeping in mind that the magnetic field interferes the motion in the $ x-y $ plane, we reduce the equation of motion to $ x-y $ plane introducing some variables. Further we will consider few bath models to analyse and to obtain time dependent orbital magnetic moment.
The solution to Eq. (\ref{q1})\cite{fordtwo, Ford1987} is:
\begin{eqnarray}
\xi(t)&=&\xi(0)\dot{G}(t)+\dot{\xi}(0)G(t)+ \int_{0}^{t}dt'G(t-t')\frac{F(t')}{m}\label{q3}
\end{eqnarray}
where $G$ is the retarded Green's function\cite{fordPhysicaa}. 
We have set $ \xi(0)=0 $.
The retarded Green's function is given by,
\begin{eqnarray}
G(t)&=&\frac{1}{2\pi}\int^{\infty}_{-\infty}d\omega e^{-i\omega t}\alpha(\omega)\label{q5}
\end{eqnarray}
where,
\begin{eqnarray}
\alpha(\omega)&=&\frac{1}{-\omega^{2}-i\omega K(\omega)+\omega_{0}^{2}+\omega\omega_{c}}\label{q6}
\end{eqnarray}\\
is the response function.
Here, $\omega_{c}=\frac{qB}{mc}$ is the cyclotron frequency.
Notice that the retarded Green's function, is a causal function, i.e.,
\begin{eqnarray}
G(t)=0 \;\;\text{for}\;\; t\leq 0
\end{eqnarray}
Correspondingly the response function $ \alpha(\omega) $ is analytic in the upper half complex plane. In other words, it does not have any poles in the upper half complex plane\cite{fordPhysicaa}.
\section{Time dependent Orbital Magnetic moment in the absence of a potential}
The time dependent orbital magnetic moment is defined as \cite{marathe, Dattagupta}:
\begin{eqnarray}
M_{z}(t) &=& \frac{-q}{2c}\langle\vec{\dot{r}}\times\vec{r} \rangle\\
&=& \frac{-q}{2c}\langle \left(\dot{x} y-\dot{y}x \right) \rangle \label{q7}
\end{eqnarray}
Using Eq. (\ref{q2}), one can write:
\begin{eqnarray}
M_{z}(t) &=& \frac{q}{2c}\mathrm{Im}\langle \dot{\xi}(t) \xi^{\dagger}(t) \rangle \label{q8}
\end{eqnarray}
The expression for $ M_{z}(t) $ can be obtained using Eq. (\ref{q3}). Here $ \xi^{\dagger}(t)$ is the Hermitian conjugate of $ \xi(t)$. 
Using a particular heat bath model, we can obtain an explicit expression for the response function and hence analyse the orbital magnetic moment. In this paper we derive the solution (Eq. (\ref{q3})) for an Ohmic bath.
The form of the kernel for the Ohmic 
dissipation model \cite{Leggett, fordtwo, ford3} is $ K(t)=2\gamma \delta(t) $. Thus, $ K(\omega)=\gamma$. 

The response function for the Ohmic bath in the absence of a confining potential i.e. $ \omega_{0}=0 $ is given by
\begin{eqnarray}
\alpha(\omega)&=&\frac{1}{-\omega^{2}-i\omega\gamma +\omega\omega_{c}}\label{q9}
\end{eqnarray}
The Green's function is then the inverse Fourier transform of $ \alpha(\omega) $ :
\begin{eqnarray}
G(t)&=&\frac{1}{\overline{\gamma}}\left(1-e^{-\overline{\gamma}t} \right) \label{q10}
\end{eqnarray}
where, $ \overline{\gamma}=\gamma + i\omega_{c} $.
The solution to Eq. (\ref{q1}) i.e. Eq. (\ref{q3}) is given by
\begin{eqnarray}
\xi(t)&=&\frac{\dot{\xi}(0)\left\lbrace 1-e^{-\overline{\gamma}t} \right\rbrace}{\overline{\gamma}} +\frac{1}{\overline{\gamma}}\int_{0}^{t}dt' \left(1- e^{-\overline{\gamma}(t-t')}\right)\nonumber\\
&&\frac{ F(t')}{m}\label{q11}
\end{eqnarray}
and 
\begin{eqnarray}
\dot{\xi}(t)&=&\dot{\xi}(0)e^{-\overline{\gamma}t}+\int_{0}^{t}dt'e^{-\overline{\gamma}(t-t')}\frac{ F(t')}{m}\label{q12}\\
\xi^{\dagger}(t)&=&\frac{\dot{\xi}^{\dagger}(0)\left\lbrace 1-e^{-\overline{\gamma}^{\dagger}t} \right\rbrace}{\overline{\gamma}^{\dagger}} +\frac{1}{\overline{\gamma}^{\dagger}}\int_{0}^{t}dt' \left(1- e^{-\overline{\gamma}^{\dagger}(t-t')}\right)\nonumber\\
&&\frac{ F^{\dagger}(t')}{m}\label{q12a}
\end{eqnarray}
where $ \overline{\gamma}^{\dagger} $ is the complex conjugate of $ \overline{\gamma} $.
Thus,
\begin{eqnarray}
M_{z}(t) &=&\frac{q}{2c}\mathrm{Im}\Biggl\lbrace \frac{\langle\vert \dot{\xi}(0) \vert^2\rangle}{\overline{\gamma}^{\dagger}} e^{-\overline{\gamma}t} 
\left(1-e^{-\overline{\gamma}^{\dagger}t} \right)   \nonumber\\
&+&\frac{\gamma\hbar}{\pi m}\int_{-\infty}^{\infty}\frac{d\omega}{\overline{\gamma}^{\dagger}}\frac{\omega}{\gamma^{2}+(\omega-\omega_{c})^{2}}\mathrm{coth}\left( \frac{\hbar\omega}{2k_{B}T}\right)\nonumber\\
&&\left( e^{-i\omega t}-e^{-\overline{\gamma}t}\right) \left( e^{-\overline{\gamma}^{\dagger}t}-1\right) \Biggl.\nonumber\\ 
&+&\frac{\gamma\hbar}{\pi m}\int_{-\infty}^{\infty}\frac{d\omega}{i}\frac{1}{\gamma^{2}+(\omega-\omega_{c})^{2}}\mathrm{coth}\left( \frac{\hbar\omega}{2k_{B}T}\right)\nonumber\\
&&\left( e^{-i\omega t}-e^{-\overline{\gamma}t}\right) \left( e^{i\omega t}-1\right)\Biggl.\Biggl\rbrace \label{q13}
\end{eqnarray}
The explicit expression for $ \langle\vert \dot{\xi}(0) \vert^2\rangle $ has been calculated in the Appendix. For an Ohmic bath, in the absence of a  potential, using Eq. (\ref{a3}) we get:
\begin{eqnarray}
\langle\vert \dot{\xi}(0) \vert^2\rangle &=&\frac{\hbar}{m\pi}\int_{-\infty}^{\infty}\frac{\omega \gamma \mathrm{coth}\left( \frac{\hbar\omega}{2k_{B}T}\right)d\omega}{\gamma^2 +\left(\omega-\omega_{c}\right)^2}\label{q14}
\end{eqnarray}
Therefore, the magnetic moment is
\begin{eqnarray}
M_{z}(t)
%&=& \frac{q}{2c}\mathrm{Im}\Biggl\lbrace \frac{\gamma\hbar}{m\pi}\int_{-\infty}^{\infty}d\omega\frac{\omega  \mathrm{coth}\left( \frac{\hbar\omega}{2k_{B}T}\right)e^{-\overline{\gamma}t} 
%\left(1-e^{-\overline{\gamma}^{\dagger}t}\right)}{\overline{\gamma}^{\dagger}\left[\gamma^2 +\left(\omega-\omega_{c}\right)^2\right]}\nonumber\\
%&+&\frac{\gamma\hbar}{\pi m}\int_{-\infty}^{\infty}d\omega\frac{\omega\mathrm{coth}\left( \frac{\hbar\omega}{2k_{B}T}\right)\left( e^{-i\omega t}-e^{-\overline{\gamma}t}\right) \left( e^{-\overline{\gamma}^{\dagger}t}-1\right)}{\overline{\gamma}^{\dagger}\left[\gamma^{2}+(\omega-\omega_{c})^{2}\right]}
%\Biggl.\nonumber\\ 
%&+&\frac{\gamma\hbar}{\pi m}\int_{-\infty}^{\infty}\frac{d\omega}{i}\frac{\mathrm{coth}\left( \frac{\hbar\omega}{2k_{B}T}\right)}{\gamma^{2}+(\omega-\omega_{c})^{2}}\left( e^{-i\omega t}-e^{-\overline{\gamma}t}\right) \left( e^{i\omega t}-1\right)\Biggl.\Biggl\rbrace \nonumber\\
&=& \frac{q\gamma\hbar}{2cm\pi}\mathrm{Im}\Biggl\lbrace \int_{-\infty}^{\infty}d\omega\omega\mathrm{coth}\left( \frac{\hbar\omega}{2k_{B}T}\right)\Biggl.\nonumber\\
&&\Bigg[\frac{1-e^{-i\omega t}-e^{-\overline{\gamma}t}e^{i\omega t}+e^{-\overline{\gamma}t}}{\omega \overline{\gamma}^{\dagger}\left(\omega +i\overline{\gamma} \right)}\Bigg.\nonumber\\
&-&\frac{1-e^{-\overline{\gamma}^{\dagger}t}e^{-i\omega t}-e^{-\overline{\gamma}t}e^{i\omega t}+e^{-\overline{\gamma}t}e^{-\overline{\gamma}^{\dagger}t}}{\overline{\gamma}^{\dagger}\left(\omega +i\overline{\gamma} \right)\left(\omega -i\overline{\gamma}^{\dagger} \right)}\nonumber\\
&+& \Bigg.\Biggl.\frac{e^{-\overline{\gamma}t} 
\left(1-e^{-\overline{\gamma}^{\dagger}t}\right)}{\overline{\gamma}^{\dagger}\left(\omega +i\overline{\gamma} \right)\left(\omega -i\overline{\gamma}^{\dagger} \right)}
\Bigg]\Biggl\rbrace \label{q15}  
\end{eqnarray}
\subsection{The High Temperature domain}
%\begin{figure}
%\includegraphics[scale=0.36]{fig1paper3.pdf}
%\caption{\label{fig1}
%Plot of the time dependent orbital magnetic moment as a function time under the condition, $ \gamma>>\omega_{c} $. This is the high temperature regime and potential is zero. 
%Here, $ \gamma=10, \omega_c=0.001,T=10 $.
%}
%\end{figure}
\begin{figure}
\includegraphics[scale=0.36]{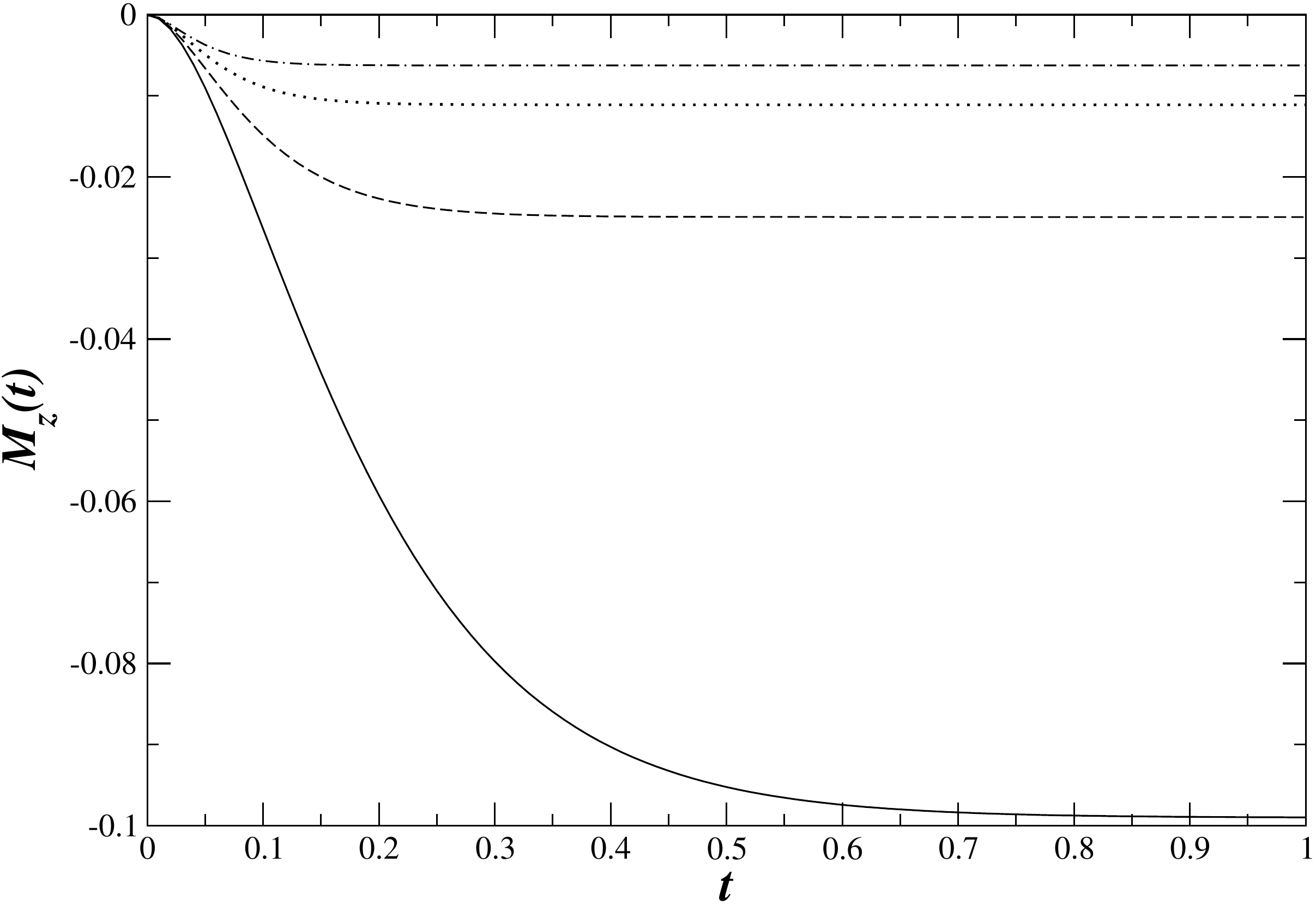}
\caption{\label{fig1}
%Plot of the time dependent orbital magnetic moment as a function of time in arbitrary units ....' 
Plot of the time dependent orbital magnetic moment as a function of time under the condition, $ \gamma>\omega_{c} $. The different curves are obtained for different values of $ \gamma $, using the high temperature condition. The solid curve is for $ \gamma=10 $, the dashed curve is for $\gamma=20  $, the dotted curve is for $\gamma=30  $ and the dot-dashed curve is for $ \gamma=40 $. For all the curves the cyclotron frequency $  \omega_{c}=1$. The orbital magnetic moment has been displayed in units of $ \frac{q\hbar}{mc} $ and the time $ t $ has been displayed in units of $ \gamma^{-1} $.
%Here, $ \omega_c=1,T=10 $.
}
\end{figure}
\begin{figure}
\includegraphics[scale=0.36]{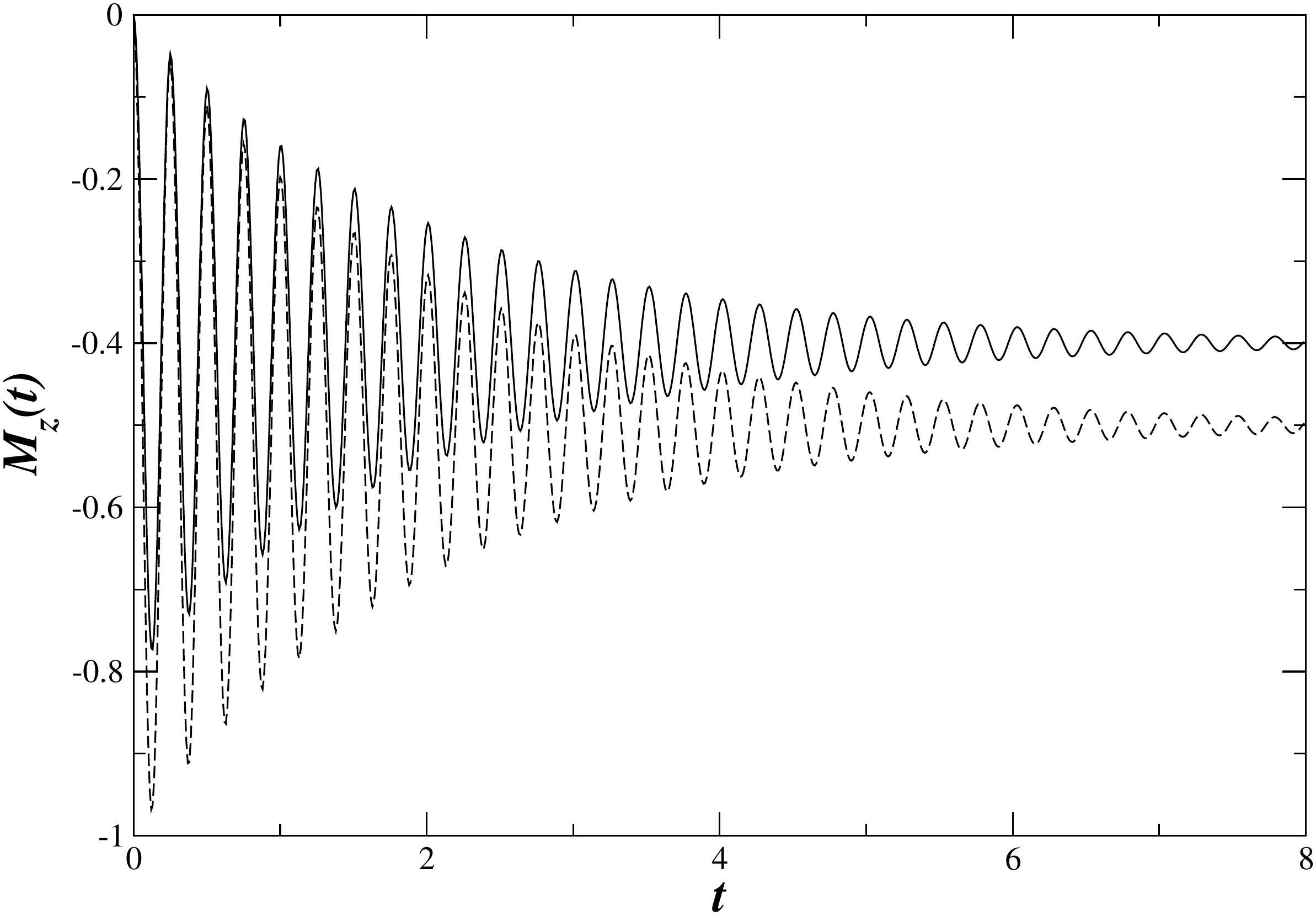}
\caption{\label{fig2}
Plot of the time dependent orbital magnetic moment as a function of time under the condition, $ \gamma<\omega_{c} $. The solid curve corresponds to the high temperature condition and the dashed curve corresponds to the low temperature condition. Both curves are obtained in the absence of a confining potential. The orbital magnetic moment has been displayed in units of $ \frac{q\hbar}{mc} $ and the time $ t $ has been displayed in units of $ \gamma^{-1} $.
%Here, $ \gamma=0.5, \omega_c=25,T=10 $.
}
\end{figure} 
In the high temperature limit $ \mathrm{coth}\left( \frac{\hbar\omega}{2k_{B}T}\right) \sim \frac{2k_{B}T}{\hbar\omega} $, thus Eq. (\ref{q15}) reduces to,
\begin{eqnarray}
M_{z}(t)&=&\frac{q\gamma k_{B}T}{cm\pi}\mathrm{Im}\Biggl\lbrace \int_{-\infty}^{\infty}d\omega
\Biggl.\nonumber\\
&&\Bigg[\frac{1-e^{-i\omega t}-e^{-\overline{\gamma}t}e^{i\omega t}+e^{-\overline{\gamma}t}}{\omega \overline{\gamma}^{\dagger}\left(\omega +i\overline{\gamma} \right)}\Bigg.\nonumber\\
&-&\frac{1-e^{-\overline{\gamma}^{\dagger}t}e^{-i\omega t}-e^{-\overline{\gamma}t}e^{i\omega t}+e^{-\overline{\gamma}t}e^{-\overline{\gamma}^{\dagger}t}}{\overline{\gamma}^{\dagger}\left(\omega +i\overline{\gamma} \right)\left(\omega -i\overline{\gamma}^{\dagger} \right)}\nonumber\\
&+& \Bigg.\Biggl.\frac{e^{-\overline{\gamma}t} 
\left(1-e^{-\overline{\gamma}^{\dagger}t}\right)}{\overline{\gamma}^{\dagger}\left(\omega +i\overline{\gamma} \right)\left(\omega -i\overline{\gamma}^{\dagger} \right)}
\Bigg]\Biggl\rbrace \label{q16}
\end{eqnarray}
Using Cauchy's residue theorem the integrals in Eq. (\ref{q16}) can be solved and the time dependent magnetic moment reduces to
\begin{eqnarray}
M_{z}(t)&=&\frac{q k_{B}T}{cm}\left\lbrace \frac{\left(\omega_c \mathrm{cos}(\omega_c t)+{\gamma}\mathrm{sin}(\omega_c t)\right)e^{-{\gamma}t}}{\gamma^2 +\omega_{c}^{2}}\right.\nonumber\\
&-&\left.
\frac{\omega_c}{\gamma^2 +\omega_{c}^{2}}\right\rbrace  \label{q19} 
\end{eqnarray}
Let us first consider the viscous damping dominated ($\gamma>\omega_{c}$) regime. The plot of the magnetic moment as a function of time for this regime is shown in Fig. (\ref{fig1}). The different curves are obtained for different values of $ \gamma $. It can be seen that as $ \gamma $ increases i.e. from the solid curve to the dot-dashed curve, the magnetic moment tends to the Bohr Van Leeuwen limit of a zero magnetic moment. In fact, for $ \gamma t>> \omega_c t $ in the limit $ \gamma t\longrightarrow \infty $, $ M_z(t)\longrightarrow 0 $.
The plot of the magnetic moment as a function of time for $\gamma<\omega_{c}$ is shown by the solid curve in Fig. (\ref{fig2}). This is the  magnetic field dominated regime. The magnetic moment shows a damped oscillatory behaviour. 
\subsection{The Low Temperature domain}
%\begin{figure}
%\includegraphics[scale=0.36]{fig3paper3.pdf}
%\caption{\label{fig3}
%Plot of the time dependent orbital magnetic moment as a function time under the condition, $ \gamma>>\omega_{c} $. This is the low temperature regime and potential is zero.
%Here, $ \gamma=10, \omega_c=0.001 $.
%}
%\end{figure}
%\begin{figure}[h]
%\includegraphics[scale=0.36]{fig4paper3.pdf}
%\caption{\label{fig4}
%Plot of the time dependent orbital magnetic moment as a function time under the condition, $ \gamma<<\omega_{c} $. This is the low temperature regime and potential is zero.
%Here, $ \gamma=0.1, \omega_c=10 $.
%}
%\end{figure}
In the low temperature limit, $ \mathrm{coth}\left( \frac{\hbar\omega}{2k_{B}T}\right) \sim 1 $, thus Eq. (\ref{q15}) reduces to,
\begin{eqnarray}
M_{z}(t)&=& \frac{q\gamma\hbar}{2cm\pi}\mathrm{Im}\Biggl\lbrace \int_{-\infty}^{\infty}d\omega\omega
\Biggl.\nonumber\\
&&\Bigg[\frac{1-e^{-i\omega t}-e^{-\overline{\gamma}t}e^{i\omega t}+e^{-\overline{\gamma}t}}{\omega \overline{\gamma}^{\dagger}\left(\omega +i\overline{\gamma} \right)}\Bigg.\nonumber\\
&-&\frac{1-e^{-\overline{\gamma}^{\dagger}t}e^{-i\omega t}-e^{-\overline{\gamma}t}e^{i\omega t}+e^{-\overline{\gamma}t}e^{-\overline{\gamma}^{\dagger}t}}{\overline{\gamma}^{\dagger}\left(\omega +i\overline{\gamma} \right)\left(\omega -i\overline{\gamma}^{\dagger} \right)}\nonumber\\
&+& \Bigg.\Biggl.\frac{e^{-\overline{\gamma}t} 
\left(1-e^{-\overline{\gamma}^{\dagger}t}\right)}{\overline{\gamma}^{\dagger}\left(\omega +i\overline{\gamma} \right)\left(\omega -i\overline{\gamma}^{\dagger} \right)}
\Bigg]\Biggl\rbrace \label{eq17}
\end{eqnarray}
Solving the above integrals using Cauchy's residue theorem, Eq. (\ref{eq17}) reduces to
\begin{eqnarray}
M_{z}(t)&=&\frac{-q \hbar}{2cm \left(\gamma^2 +\omega_{c}^2 \right)}\left\lbrace \left(\gamma^2 +\omega_{c}^2 \right)+ 2\omega_{c}\gamma e^{-\gamma t}\mathrm{sin}(\omega_{c}t)\right.\nonumber\\
&+&\left.\left(\gamma^2 -\omega_{c}^2 \right)e^{-\gamma t}\mathrm{cos}(\omega_{c}t)-2\gamma^2 e^{-2\gamma t}  \right\rbrace \label{eq18}
\end{eqnarray}
%\subsubsection{$\gamma>>\omega_{c}$}
%Under this condition, Eq. (\ref{eq18}) reduces to,
%\begin{eqnarray}
%M_{z}(t)&=&\frac{-q \hbar}{2cm}
%\end{eqnarray}
\begin{figure}[t]
\includegraphics[scale=0.36]{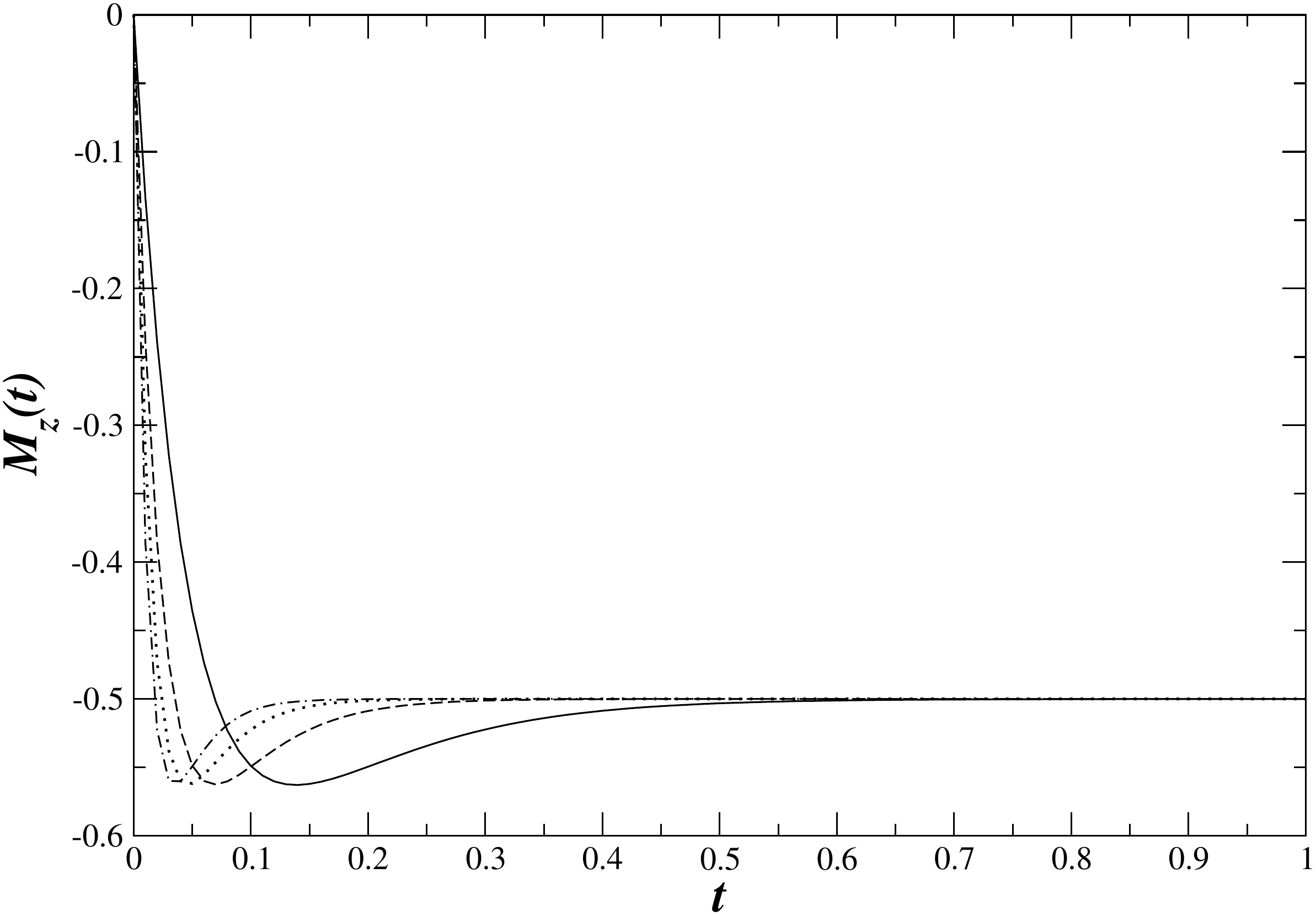}
\caption{\label{fig3}
Plot of the time dependent orbital magnetic moment as a function of time under the condition, $ \gamma>\omega_{c} $. The different curves are obtained for different values of $ \gamma $, using the low temperature condition. The solid curve is for $ \gamma=10 $, the dashed curve is for $\gamma=20  $, the dotted curve is for $\gamma=30  $ and the dot-dashed curve is for $ \gamma=40 $. For all the curves the cyclotron frequency $  \omega_{c}=1$. The orbital magnetic moment has been displayed in units of $ \frac{q\hbar}{mc} $ and the time $ t $ has been displayed in units of $ \gamma^{-1} $.
%Here, $ \omega_c=1,T=10 $.
}
\end{figure}
In the damping dominated ($ \gamma>\omega_{c} $) regime at low temperatures, we notice that the magnetic moment goes over to the constant value $ -\frac{q\hbar}{2mc} $ in the limit $ t\longrightarrow \infty $ exhibiting diamagnetism.
We display the magnetic moment $ M_z(t) $ as a function of time $ t $ in Fig. (\ref{fig3}) for various strengths of $ \gamma $ in the low temperature domain for $ \gamma > \omega_c $. We notice that all curves settle down to the constant value of $ -\frac{q\hbar}{2mc} $ in the $ t\longrightarrow \infty $ limit. As $ \gamma $ increases it settles down to the constant negative value at a smaller value of $ t $. In contrast to the high temperature zero magnetic moment asymptotic time limit (see Fig. (\ref{fig1})), in the low temperature domain the magnetic moment settles down to a constant negative value of $ -\frac{q\hbar}{2mc} $ for all values of $ \gamma $ exhibiting diamagnetism. 
%Here also the different curves are magnetic moment for different values of $ \gamma $. The higher the value of $ \gamma $, the lesser time is required for the magnetic moment to settle to the constant value. The low temperature behaviour in contrast to the high temperature damping dominating regime where the orbital magnetic moment goes to zero in the limit $ \gamma t\longrightarrow \infty $, in accordance with the Bohr Van Leeuwen theorem.
%\subsubsection{$\gamma<<\omega_{c}$}
%Under this condition, Eq. (\ref{eq18}) reduces to,
%\begin{eqnarray}
%M_{z}(t)&=&\frac{q \hbar}{4cm}\left\lbrace 2\mathrm{cos(\omega_c t)}-2  \right\rbrace
%\end{eqnarray}
The plot of the magnetic moment as a function of time for $ \gamma<\omega_{c} $ is shown by the dashed curve in Fig. (\ref{fig2}). As in the high temperature regime (the solid curve in Fig. (\ref{fig2})) we again notice a damped oscillatory behaviour of the time dependent orbital magnetic moment in the low temperature regime (the dashed curve in Fig. (\ref{fig2})). However, as expected, we do see a quantitative difference between the low temperature (dashed) curve and the high temperature (solid) curve.

\section{Time dependent Orbital Magnetic moment in the presence of a potential}
The response function in the presence of a confining harmonic oscillator potential is given by
\begin{eqnarray}
\alpha(\omega)&=&\frac{1}{-\omega^{2}-i\omega\gamma +\omega\omega_{c}+\omega_{0}^{2}}
\end{eqnarray}
The Green's function is then the inverse Fourier transform of $ \alpha(\omega) $ 
\begin{eqnarray}
G(t)&=&\frac{1}{\gamma_{+}-\gamma_{-}} \left( e^{\gamma_{+}t}-e^{\gamma_{-}t}\right) 
\end{eqnarray}
where, $$ \gamma_{\pm}=\frac{-\overline{\gamma}}{2}\pm\frac{1}{2}\sqrt{\overline{\gamma}^2-4\omega_{0}^{2}} $$
The solution to Eq. (\ref{q1}) i.e. Eq. (\ref{q3}) is given by
\begin{eqnarray}
\xi(t)&=&\frac{\dot{\xi}(0)}{\gamma_{+}-\gamma_{-}}\left\lbrace e^{\gamma_{+}t}-e^{\gamma_{-}t}\right\rbrace \nonumber\\
&+& \int_{0}^{t}d\tau \frac{\left\lbrace e^{\gamma_{+}(t-\tau)}-e^{\gamma_{-}(t-\tau)} \right\rbrace}{\gamma_{+}-\gamma_{-}} \frac{F(\tau)}{m}\label{e11}
\end{eqnarray}
and 
\begin{eqnarray}
\dot{\xi}(t)&=&\frac{\dot{\xi}(0)}{\gamma_{+}-\gamma_{-}}\left\lbrace \gamma_{+}e^{\gamma_{+}t}-\gamma_{-}e^{\gamma_{-}t}\right\rbrace \nonumber\\
&+& \int_{0}^{t}d\tau \frac{\left\lbrace \gamma_{+}e^{\gamma_{+}(t-\tau)}-\gamma_{-}e^{\gamma_{-}(t-\tau)} \right\rbrace}{\gamma_{+}-\gamma_{-}} \frac{F(\tau)}{m}\label{e12}\\
\xi^{\dagger}(t)&=&\frac{\dot{\xi}^{\dagger}(0)}{\gamma_{+}^{*}-\gamma_{-}^{*}}\left\lbrace e^{\gamma_{+}^{*}t}-e^{\gamma_{-}^{*}t}\right\rbrace \nonumber\\
&+& \int_{0}^{t}d\tau \frac{\left\lbrace e^{\gamma_{+}^{*}(t-\tau)}-e^{\gamma_{-}^{*}(t-\tau)} \right\rbrace}{\gamma_{+}^{*}-\gamma_{-}^{*}} \frac{F^{\dagger}(\tau)}{m}\label{e12a}
\end{eqnarray}
where, $ \gamma_{\pm}^{*} $ are the complex conjugates of $ \gamma_{\pm} $.
Using Eqs. (\ref{e12}) and (\ref{e12a}), one can obtain the orbital magnetic moment
\begin{eqnarray}
M_{z}(t)&=&\frac{q}{2c}\mathrm{Im}\Biggl\lbrace \frac{\langle\vert \dot{\xi}(0) \vert^2\rangle\left(e^{\gamma_{+}^{*}t}-e^{\gamma_{-}^{*}t} \right)\left(\gamma_{+}e^{\gamma_{+}t}-\gamma_{-}e^{\gamma_{-}t} \right)}{\vert \gamma_{+}-\gamma_{-} \vert^{2}}\Biggl.\nonumber\\
&+&\frac{\gamma\hbar}{\pi m}\int_{-\infty}^{\infty}d\omega \frac{\omega}{\vert \gamma_{+}-\gamma_{-} \vert^{2}}\mathrm{coth} \left( \frac{\hbar\omega}{2k_{B}T}\right) \nonumber\\
&&\hspace*{-0.3cm}\left[\frac{e^{i\omega t}}{\omega +i\gamma_{+}^{*}}-\frac{e^{i\omega t}}{\omega +i\gamma_{-}^{*}}-\frac{e^{\gamma_{+}^{*}t}}{\omega +i\gamma_{+}^{*}}+\frac{e^{\gamma_{-}^{*}t}}{\omega +i\gamma_{-}^{*}} \right]\nonumber\\
&&\hspace*{-0.3cm}
\Biggl.\Biggl[ \frac{\gamma_{+}e^{-i\omega t}}{\omega -i\gamma_{+}}-\frac{\gamma_{-}e^{-i\omega t}}{\omega -i\gamma_{-}}-\frac{\gamma_{+}e^{\gamma_{+}t}}{\omega -i\gamma_{+}}+\frac{\gamma_{-}e^{\gamma_{-}t}}{\omega -i\gamma_{-}}\Biggl]   
\Biggl\rbrace \label{e13}
\end{eqnarray}
In this case, using Eq. (\ref{a3}) $ \langle\vert \dot{\xi}(0) \vert^2\rangle $ is given by
\begin{eqnarray}
\langle\vert \dot{\xi}(0) \vert^2\rangle &=&\frac{\hbar}{m\pi}\int_{-\infty}^{\infty}\frac{\omega^3 \gamma\mathrm{coth}\left( \frac{\hbar\omega}{2k_{B}T}\right)d\omega}{\left(\omega^{2}-\omega_{0}^{2}-\omega\omega_{c}\right)^2+\left(\omega \gamma\right)^2}
\end{eqnarray}
Thus Eq. (\ref{e13}) reduces to
\begin{eqnarray}
M_{z}(t)&=&\frac{q}{2c}\mathrm{Im}\Biggl\lbrace \frac{\hbar}{m\pi}\frac{1}{\vert \gamma_{+}-\gamma_{-} \vert^{2}}\Biggl.\nonumber\\
&&\int_{-\infty}^{\infty}\frac{\omega^3 \gamma\mathrm{coth}\left( \frac{\hbar\omega}{2k_{B}T}\right)d\omega}{\left(\omega^{2}-\omega_{0}^{2}-\omega\omega_{c}\right)^2+\left(\omega \gamma\right)^2}\nonumber\\
&&\left(e^{\gamma_{+}^{*}t}-e^{\gamma_{-}^{*}t} \right)\left(\gamma_{+}e^{\gamma_{+}t}-\gamma_{-}e^{\gamma_{-}t} \right)\nonumber\\
&+&\frac{\gamma\hbar}{\pi m}\int_{-\infty}^{\infty}d\omega \frac{\omega}{\vert \gamma_{+}-\gamma_{-} \vert^{2}}\mathrm{coth} \left( \frac{\hbar\omega}{2k_{B}T}\right)\nonumber\\
&&\hspace*{-1cm}\left[\frac{e^{i\omega t}}{\omega +i\gamma_{+}^{*}}-\frac{e^{i\omega t}}{\omega +i\gamma_{-}^{*}}-\frac{e^{\gamma_{+}^{*}t}}{\omega +i\gamma_{+}^{*}}+\frac{e^{\gamma_{-}^{*}t}}{\omega +i\gamma_{-}^{*}} \right] \nonumber\\
&&\hspace*{-1.1cm}\Biggl.\Biggl[\frac{\gamma_{+}e^{-i\omega t}}{\omega -i\gamma_{+}}-\frac{\gamma_{-}e^{-i\omega t}}{\omega -i\gamma_{-}}-\frac{\gamma_{+}e^{\gamma_{+}t}}{\omega -i\gamma_{+}}+\frac{\gamma_{-}e^{\gamma_{-}t}}{\omega -i\gamma_{-}}\Biggl]   
\Biggl\rbrace\label{e13a}
\end{eqnarray}
\subsection{The High Temperature domain}
\begin{figure}
\includegraphics[scale=0.36]{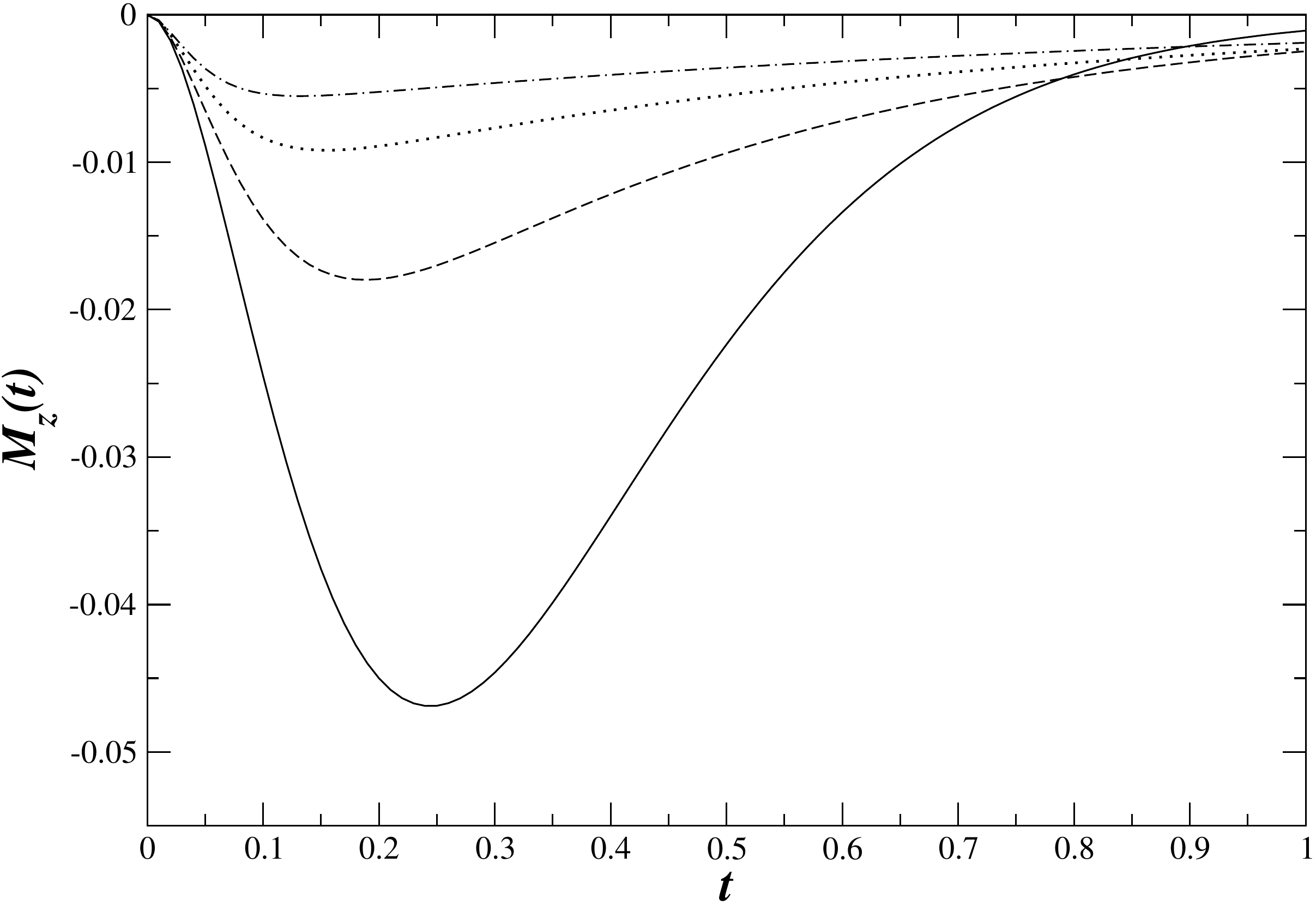}
\caption{\label{fig4}
Plot of the time dependent orbital magnetic moment as a function of time under the condition, $ \gamma>\omega_{c} $. The different curves are obtained for different values of $ \gamma $, using the high temperature condition. The solid curve is for $ \gamma=10 $, the dashed curve is for $\gamma=20  $, the dotted curve is for $\gamma=30  $ and the dot-dashed curve is for $ \gamma=40 $. For all the curves the cyclotron frequency $  \omega_{c}=1$ and $ \omega_0=5 $. The curves are obtained in the presence of a finite potential. The orbital magnetic moment has been displayed in units of $ \frac{q\hbar}{mc} $ and the time $ t $ has been displayed in units of $ \gamma^{-1} $.
%Here, $ \omega_0=1,T=10 $.
}
\end{figure}
\begin{figure}
\includegraphics[scale=0.36]{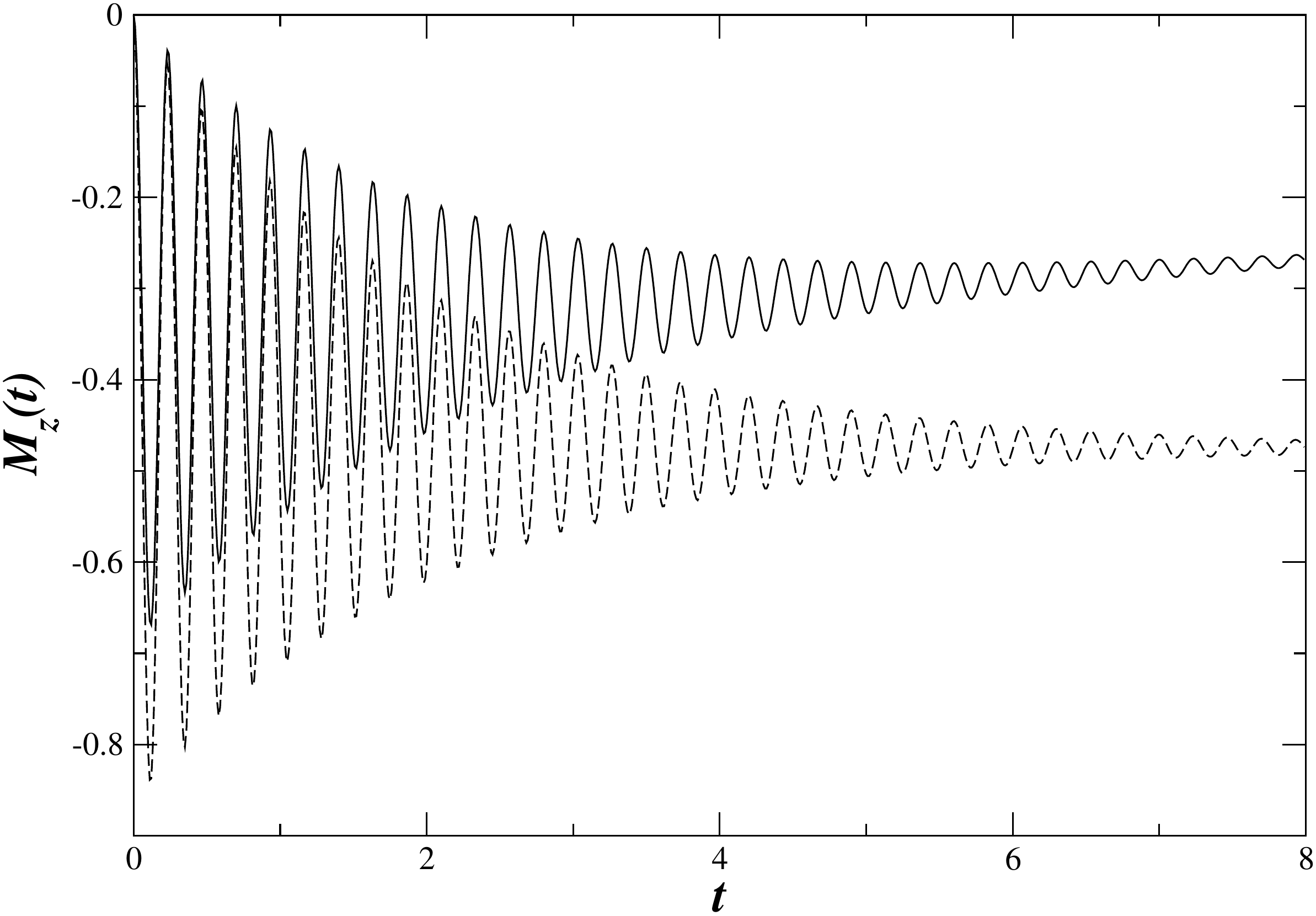}
\caption{\label{fig7}
Plot of the time dependent orbital magnetic moment as a function time under the condition, $ \gamma<\omega_{c} $. The solid curve corresponds to the high temperature condition and the dashed curve corresponds to the low temperature condition. Both curves are obtained in the presence of a finite potential. The orbital magnetic moment has been displayed in units of $ \frac{q\hbar}{mc} $ and the time $ t $ has been displayed in units of $ \gamma^{-1} $. Here, $ \gamma=0.5, \omega_{c}=25$ and $ \omega_0=5 $.
}
\end{figure}
In the high temperature limit $ \mathrm{coth}\left( \frac{\hbar\omega}{2k_{B}T}\right) \sim \frac{2k_{B}T}{\hbar\omega} $. Substituting this in Eq. (\ref{e13a}) and using Cauchy's residue theorem, the magnetic moment in the high temperature limit can be obtained:
\begin{eqnarray}
M_z(t)&=& -\frac{2qk_B T\gamma}{mc\vert \gamma_{+}-\gamma_{-} \vert^{2}}\mathrm{Im}\Biggl\lbrace
\frac{\gamma_{+}}{(\gamma_{+}+\gamma_{+}^{*})}
-\frac{\gamma_{+}e^{\gamma_{+}t}e^{\gamma_{+}^{*}t}}{(\gamma_{+}+\gamma_{+}^{*})}\Biggl.\nonumber\\
&-&\frac{\gamma_{-}}{(\gamma_{-}+\gamma_{+}^{*})}
+\frac{\gamma_{-}e^{\gamma_{-}t}e^{\gamma_{+}^{*}t}}{(\gamma_{-}+\gamma_{+}^{*})}
-\frac{\gamma_{+}}{(\gamma_{+}+\gamma_{-}^{*})}\nonumber\\
&+&\frac{\gamma_{+}e^{\gamma_{+}t}e^{\gamma_{-}^{*}t}}{(\gamma_{+}+\gamma_{-}^{*})}+\frac{\gamma_{-}}{(\gamma_{-}+\gamma_{-}^{*})}-\frac{\gamma_{-}e^{\gamma_{-}t}e^{\gamma_{-}^{*}t}}{(\gamma_{-}+\gamma_{-}^{*})}\nonumber\\
&+&\frac{\gamma_{+}^{2}\left(e^{\gamma_{+}^{*}t}-e^{\gamma_{-}^{*}t} \right)\left(\gamma_{+}e^{\gamma_{+}t}-\gamma_{-}e^{\gamma_{-}t} \right)}{(\gamma_{+}-\gamma_{-})(\gamma_{+}+\gamma_{+}^{*})(\gamma_{+}+\gamma_{-}^{*})}\nonumber\\
&+&
\frac{\gamma_{-}^{2}\left(e^{\gamma_{+}^{*}t}-e^{\gamma_{-}^{*}t}\right)
\left(\gamma_{+}e^{\gamma_{+}t}-\gamma_{-}e^{\gamma_{-}t}\right)}{(\gamma_{-}-\gamma_{+})(\gamma_{-}+\gamma_{+}^{*})(\gamma_{-}+\gamma_{-}^{*})}
\Biggl.\Biggl\rbrace \label{e14}
\end{eqnarray}
The plot of the magnetic moment as a function of time for $ \gamma>\omega_c $ regime is shown in Fig. (\ref{fig4}). The different curves are obtained for different values of $ \gamma $. It can be seen that just like in Fig. (\ref{fig1}), as $ \gamma $ increases i.e. from the solid curve to the dot-dashed curve, the magnetic moment tends to a zero value. We notice that the magnetic moment in the presence of a potential has a value closer to zero for a given value of $ \gamma $, compared to a situation where there is no confining potential. This signifies that adding a potential to the problem makes it easier to get the Bohr Van Leeuwen limit of a zero magnetic moment.
An evaluation of Eq. (\ref{e14}) gives $ M_z(t)=0 $ for the damping dominated ($\gamma>\omega_{c}$) regime in the high temperature domain in the limit $ \gamma t\longrightarrow \infty $. Thus the Bohr Van Leeuwen limit of a zero magnetic moment is recovered both in the absence and in the presence of a confining potential in the viscous damping dominated regime in the high temperature classical limit.
The plot of the magnetic moment as a function of time for $\gamma<\omega_{c}$ regime is shown by the solid curve in Fig. (\ref{fig7}). This is the magnetic field dominated regime. The magnetic moment shows a damped oscillatory behaviour.

\subsection{The Low Temperature domain}
\begin{figure}[ht]
\includegraphics[scale=0.36]{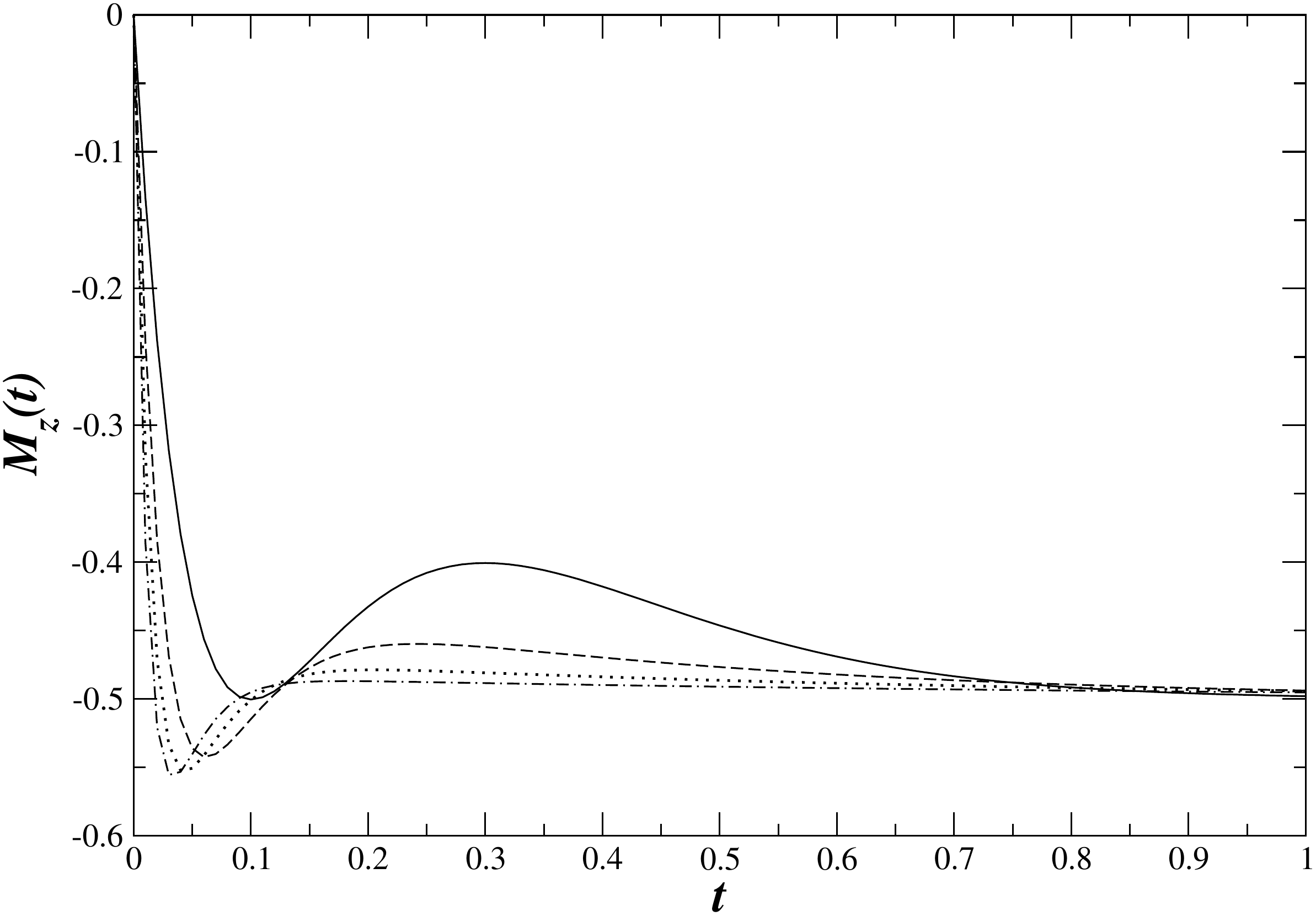}
\caption{\label{fig6}
Plot of the time dependent orbital magnetic moment as a function of time under the condition, $ \gamma>\omega_{c} $. The different curves are obtained for different values of $ \gamma $, using the low temperature condition. The solid curve is for $ \gamma=10 $, the dashed curve is for $\gamma=20  $, the dotted curve is for $\gamma=30  $ and the dot-dashed curve is for $ \gamma=40 $. For all the curves the cyclotron frequency $  \omega_{c}=1$ and $ \omega_0=5 $. The curves are obtained in the presence of a finite potential. The orbital magnetic moment has been displayed in units of $ \frac{q\hbar}{mc} $ and the time $ t $ has been displayed in units of $ \gamma^{-1} $.
%Here, $ \gamma=0.1, \omega_c=20,T=10 $.
}
\end{figure}
In the low temperature limit, $ \mathrm{coth}\left( \frac{\hbar\omega}{2k_{B}T}\right) \sim 1 $. Substituting this in Eq. (\ref{e13a}) and using Cauchy's residue theorem, the magnetic moment in the low temperature limit can be obtained:
\begin{eqnarray}
M_z(t)&=& -\frac{q\hbar\gamma}{2mc\vert \gamma_{+}-\gamma_{-} \vert^{2}}\nonumber\\
&&\mathrm{Im}\Biggl\lbrace 2i\Bigg( 
\frac{\gamma_{+}^2}{(\gamma_{+}+\gamma_{+}^{*})}
+\frac{\gamma_{+}\gamma_{+}^{*}e^{\gamma_{+}t}e^{\gamma_{+}^{*}t}}{(\gamma_{+}+\gamma_{+}^{*})}\Bigg.\Biggl.\nonumber\\
&-&\frac{\gamma_{-}^2}{(\gamma_{-}+\gamma_{+}^{*})}
-\frac{\gamma_{-}\gamma_{+}^{*}e^{\gamma_{-}t}e^{\gamma_{+}^{*}t}}{(\gamma_{-}+\gamma_{+}^{*})}
-\frac{\gamma_{+}^2}{(\gamma_{+}+\gamma_{-}^{*})}\nonumber\\
&-&\frac{\gamma_{+}\gamma_{-}^{*}e^{\gamma_{+}t}e^{\gamma_{-}^{*}t}}{(\gamma_{+}+\gamma_{-}^{*})}+\frac{\gamma_{-}^2}{(\gamma_{-}+\gamma_{-}^{*})}+\frac{\gamma_{-}\gamma_{-}^{*}e^{\gamma_{-}t}e^{\gamma_{-}^{*}t}}{(\gamma_{-}+\gamma_{-}^{*})}\nonumber\\
&-&\frac{\gamma_{+}^{3}\left(e^{\gamma_{+}^{*}t}-e^{\gamma_{-}^{*}t} \right)\left(\gamma_{+}e^{\gamma_{+}t}-\gamma_{-}e^{\gamma_{-}t} \right)}{(\gamma_{+}-\gamma_{-})(\gamma_{+}+\gamma_{+}^{*})(\gamma_{+}+\gamma_{-}^{*})}\nonumber\\
&-&\frac{\gamma_{-}^{3}\left(e^{\gamma_{+}^{*}t}-e^{\gamma_{-}^{*}t}\right)
\left(\gamma_{+}e^{\gamma_{+}t}-\gamma_{-}e^{\gamma_{-}t}\right)}{(\gamma_{-}-\gamma_{+})(\gamma_{-}+\gamma_{+}^{*})(\gamma_{-}+\gamma_{-}^{*})}
\Bigg.\Bigg) \Biggl.\Biggl\rbrace \label{e15}
\end{eqnarray}
In the viscous damping dominated regime ($\gamma>\omega_{c} $), we find that the magnetic moment goes over to a constant negative value of $ -\frac{q\hbar}{2mc} $ (see Fig. (\ref{fig6})). Here also similar to Fig. (\ref{fig3}), we find that as $ \gamma $ increases, it takes a smaller amount of time to settle down to the constant negative value.

The plot of the magnetic moment as a function of time for $ \gamma<\omega_{c} $ regime is shown by the dashed curve in Fig. (\ref{fig7}). The nature is similar to what is exhibited by the solid curve in Fig. (\ref{fig7}), which is damped oscillatory. However, in detail, the two curves (the solid line and dashed line in Fig. (\ref{fig7})) are quantitatively distinct.
\section{Approach to the high temperature classical thermodynamics}
\begin{figure}
\includegraphics[scale=0.36]{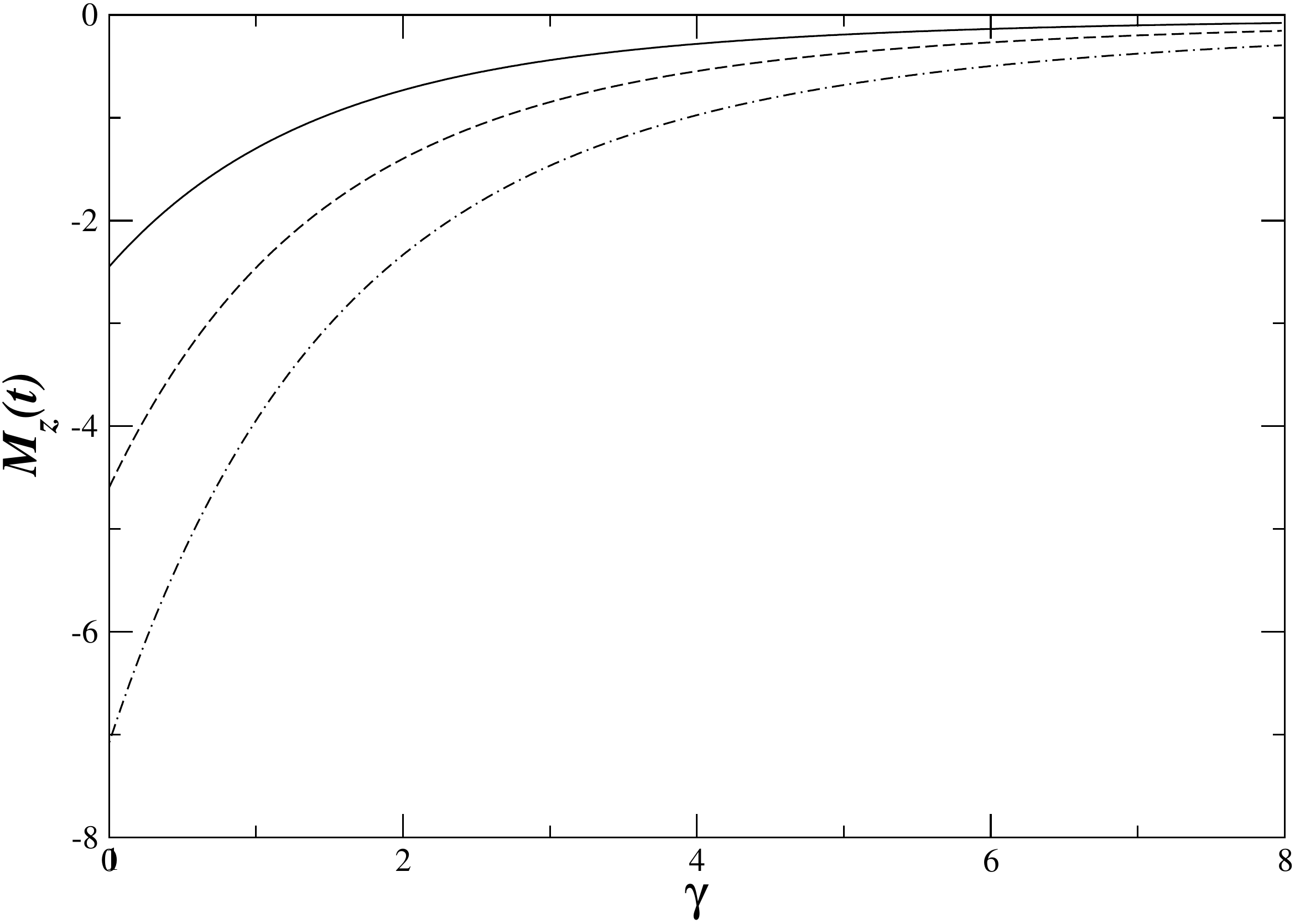}
\caption{\label{fig5}
Plot of the time dependent orbital magnetic moment as a function of the damping coefficient $ \gamma $, for different values of the cyclotron frequency $ \omega_c $ and at a finite time $ t $. This is the zero potential case. The plots are obtained using Eq. (\ref{q19}). The solid line is for $ \omega_c =0.5$, dashed is for $ \omega_c =1$ and dot-dashed is for $ \omega_c =2$. For all cases, $ t=1 $. The orbital magnetic moment has been displayed in units of $ \frac{q\hbar}{mc} $ and the time $ t $ has been scaled in units of $ \gamma^{-1} $. The damping coefficient $ \gamma $ and the cyclotron frequency $ \omega_c $ have been scaled in the units of $ \frac{k_B T}{\hbar} $. %Notice that the $ \gamma $ range has been chosen in such a way that it satisfies $ \gamma \gtrsim\omega_c $.
}
\end{figure}
\begin{figure}
\includegraphics[scale=0.36]{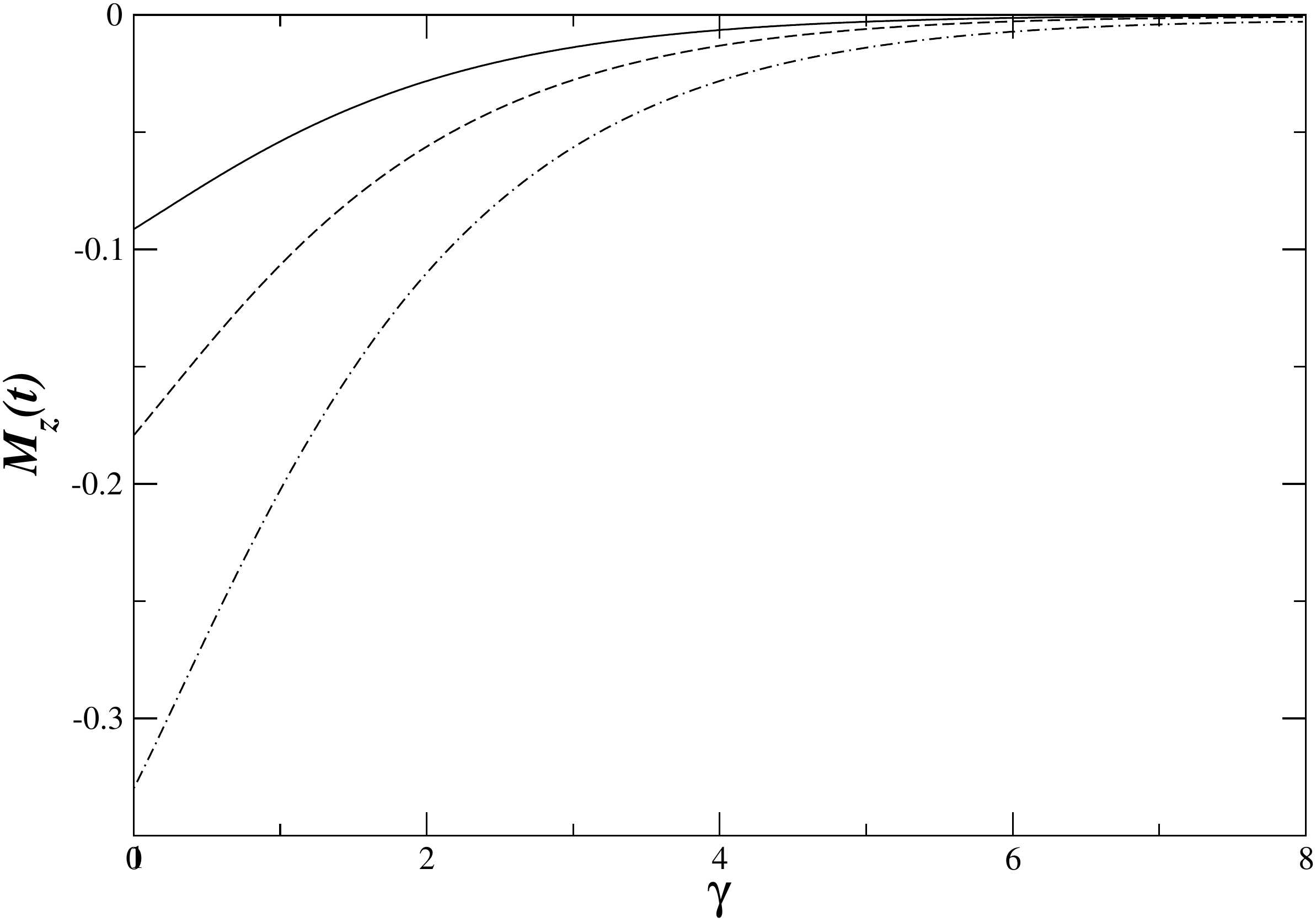}
\caption{\label{fig5new}
Plot of the time dependent orbital magnetic moment as a function of the damping coefficient $ \gamma $, for different values of the cyclotron frequency $ \omega_c $ and at a finite time $ t $. This is the finite potential case. The plots are obtained using Eq. (\ref{e14}). The solid line is for $ \omega_c =0.5$, dashed is for $ \omega_c =1$ and dot-dashed is for $ \omega_c =2$. For all the cases, $ t=1 $ and $ \omega_0 =5 $. The orbital magnetic moment has been displayed in units of $ \frac{q\hbar}{mc} $ and the time $ t $ has been scaled in units of $ \gamma^{-1} $. The damping coefficient $ \gamma $, cyclotron frequency $ \omega_c $ and the harmonic oscillator frequency $ \omega_0 $ have been scaled in the units of $ \frac{k_B T}{\hbar} $. %Notice that the $ \gamma $ range has been chosen in such a way that it satisfies $ \gamma \gtrsim\omega_c $.
}
\end{figure}
The curves of the magnetic moment as a function of $ \gamma $ for different values of $ \omega_c $ are shown in Fig. (\ref{fig5}) using Eq. (\ref{q19}) which corresponds to the high temperature damping dominated regime. It can be seen that as $ \gamma $ increases the magnetic moment tends to zero. We also notice that the lower the value of $ \omega_c $, the faster the approach to zero of the magnetic moment. This trend is similar to what has been reported in Ref.\cite{Dattaguptaprl}. In contrast to Ref.\cite{Dattaguptaprl}, where they have studied the dependence of the magnetic moment on the parameters $ \gamma $ and $ \omega_c $ in the asymptotic time limit i.e. $ t \rightarrow \infty $, we have studied this for a finite value of time $ t $. We thus get a detailed picture of the interplay of the competing time scales set by $ \gamma^{-1} $ and $ \omega_c^{-1} $ leading to a zero magnetic moment Bohr Van Leeuwen limit for a large dissipation coefficient $ \gamma $ for a finite $ t $ i.e. for $ \gamma t >>1 $.

In addition we study a similar dependence of the magnetic moment in the presence of a finite potential in Fig. (\ref{fig5new}). The plots are obtained using Eq. (\ref{e14}). In this case also we recover the Bohr Van Leeuwen limit of a zero magnetic moment. We can see if we compare Figs. (\ref{fig5}) and (\ref{fig5new}), that the magnetic moment for the same cyclotron frequency shows very different values. The magnitude of the values are much smaller in the finite potential case compared to the zero potential case.
The difference arises because of the presence of an additional time scale $ \omega_0^{-1} $. In the finite potential case the zero value is obtained for an even smaller value of $ \gamma $ compared to the case of a zero confining potential.

\section{Conclusions}
In this paper we have analysed in detail the nonequilibrium quantum Langevin dynamics of the orbital magnetic moment in the high temperature 
classical domain, as well as the low temperature quantum domain and studied the role of a confining potential in this dynamics.
For both the high temperature classical domain and low temperature quantum domain we have addressed two different regimes: a damping dominated regime $ \gamma>\omega_c $ and a magnetic field dominated regime $ \gamma<\omega_c $. We have obtained the Bohr Van Leeuwen limit in the damping dominated $ \gamma>\omega_c $ high temperature classical regime in the limit $ \gamma t \rightarrow \infty$ . This has been observed both in the absence of a potential as well as in the presence of a confining potential. In the low temperature  quantum regime, the magnetic moment settles to a constant negative value exhibiting diamagnetism. This feature remains the same whether or not the system is kept in a confining potential. We notice that the orbital magnetic moment exhibits a damped oscillatory behaviour in the magnetic field dominated regime $ \gamma<\omega_c $ both in the high temperature classical and low temperature quantum regimes, the oscillation frequency being determined by the cyclotron frequency.

Our predictions for the damped oscillatory dynamics of the orbital diamagnetic moment can be tested via cold atom experiments with hybrid traps for ions and neutral atoms, for instance, by considering a single ion immersed in a BEC\cite{one}. An uniform magnetic field can be generated using magnetic coils in a Helmholtz configuration and a dissipative medium can be set up via a 3D optical molasses\cite{three}, combined with a magnetic or an optical trap. One can tune the temperature by tuning the depth of the trap and measure the orbital magnetic moment dynamics in the entire range spanning between low and high temperatures.

We would like to draw attention to a subtlety in obtaining the Bohr Van Leeuwen zero magnetic moment limit. As we discussed, there are two significant time scale domains: one dominated by the viscous relaxation rate and the other dominated by the cyclotron frequency. We notice that in the first domain one obtains the classical Bohr Van Leeuwen limit even in the absence of a confining potential. In contrast, in the second domain, which is dominated by the cyclotron frequency, the viscous damping rate is not high enough to reach the equilibrium thermodynamic Bohr Van Leeuwen zero magnetic moment limit. Thus the relative strength of the viscous damping and the cyclotron frequency plays a crucial role in attaining the classical thermodynamic Bohr Van Leeuwen limit.

%We would like to draw attention to a subtlety in obtaining the Bohr Van Leeuwen zero magnetic moment limit. As we discussed, there are two significant time scale domains: one dominated by the viscous relaxation rate and the other dominated by the magnetic cyclotron frequency. We notice that in the first domain one obtains the classical Bohr Van Leeuwen limit even in the absence of a confining potential. In contrast, in the second domain, one needs to introduce a confining potential to recover the zero magnetic moment limit in the classical high temperature limit \cite{Dattaguptaprl}.

\section*{Acknowledgements}
We thank Michael V. Berry, Abhishek Dhar and Joseph Samuel for stimulating discussions. We would also like to thank Sanjukta Roy and Saptarishi Chaudhuri for discussions related to experimental implications of our analysis.
\renewcommand{\theequation}{A-\arabic{equation}}
  % redefine the command that creates the equation no.
  \setcounter{equation}{0}  % reset counter
\section*{Appendix: Expression for $ \langle\vert \dot{\xi}(0) \vert^2\rangle $}\label{A1}
The reduced QLE is given by Eq. (\ref{q1})
\begin{eqnarray}
\ddot{\xi}(t)&=&-\int_{0}^{t}K(t-t')\dot{\xi}(t')dt'-\frac{iqB}{mc}\dot{\xi}(t)-\omega_{0}^{2}\xi(t)\nonumber\\
&+&\frac{F(t)}{m}\label{a1}
\end{eqnarray}
In terms of Fourier transform, the solution of the above equation can be written as,
\begin{eqnarray}
\xi(\omega)&=&\frac{F(\omega)}{m\left(-\omega^{2}-i\omega K(\omega)+\omega_{0}^{2}+\omega\omega_{c}\right)}\label{a2}
\end{eqnarray}
Now,
\begin{eqnarray*}
{\xi}(t)&=&\int_{-\infty}^{\infty}\xi(\omega)e^{-i\omega t}d\omega\nonumber\\
\dot{\xi}(t)&=&\int_{-\infty}^{\infty}-i\omega\xi(\omega)e^{-i\omega t}d\omega \nonumber
\end{eqnarray*}
For $ t=0 $,
\begin{eqnarray*}
\dot{\xi}(0)&=&\int_{-\infty}^{\infty}-i\omega\xi(\omega)d\omega\nonumber\\
&=& \int_{-\infty}^{\infty}\frac{-i\omega F(\omega)d\omega}{m\left(-\omega^{2}-i\omega K(\omega)+\omega_{0}^{2}+\omega\omega_{c}\right)}\nonumber\\
&=& \int_{-\infty}^{\infty}\frac{i\omega F(\omega)d\omega}{m\left(\omega^{2}-\omega_{0}^{2}-\omega\omega_{c}-\omega\mathrm{Im}[K(\omega)]+i\omega \mathrm{Re}[K(\omega)]\right)}
\end{eqnarray*}
And,
\begin{eqnarray*}
\dot{\xi}^{\dagger}(0)&=& \int_{-\infty}^{\infty}\frac{-i\omega F^{\dagger}(\omega)d\omega}{m\left(\omega^{2}-\omega_{0}^{2}-\omega\omega_{c}-\omega\mathrm{Im}[K(\omega)]-i\omega \mathrm{Re}[K(\omega)]\right)}
\end{eqnarray*}
Hence,
\begin{eqnarray}
\langle\vert \dot{\xi}(0) \vert^2\rangle &=&\int_{-\infty}^{\infty}\omega^2 \langle F(\omega)F^{\dagger}(\omega)\rangle d\omega\nonumber\\
&&\hspace*{-0.5cm}\frac{1}{m^2\left[\left(\omega^{2}-\omega_{0}^{2}-\omega\omega_{c}-\omega\mathrm{Im}[K(\omega)]\right)^2+\left(\omega \mathrm{Re}[K(\omega)]\right)^2\right]}\nonumber\\
&=&\frac{\hbar}{m\pi}\int_{-\infty}^{\infty}\omega^3 \mathrm{Re}[K(\omega)]\mathrm{coth}\left( \frac{\hbar\omega}{2k_{B}T}\right)d\omega\nonumber\\
&&\hspace*{-0.5cm}\frac{1}{\left(\omega^{2}-\omega_{0}^{2}-\omega\omega_{c}-\omega\mathrm{Im}[K(\omega)]\right)^2+\left(\omega \mathrm{Re}[K(\omega)]\right)^2} \label{a3}
\end{eqnarray}
The last step is obtained using the force-force correlation (Eq. (\ref{e3})).

\bibliography{supurnareferences}
\end{document}